\documentclass[a4paper,aps,twocolumn,nofootinbib]{revtex4}
\RequirePackage[colorlinks,hyperindex]{hyperref}
\RequirePackage[english]{babel}
\RequirePackage[latin1]{inputenc}
\RequirePackage[T1]{fontenc}
\RequirePackage{mathrsfs}
\RequirePackage{amsmath}
\RequirePackage{amssymb}
\RequirePackage{amsbsy}
\RequirePackage{color}
\RequirePackage{bm}
\hypersetup{colorlinks=true,breaklinks=true,urlcolor=blue,linkcolor=red}
\begin{document}
\title{\bf{de Broglie-Bohm formulation of Dirac fields}}
\author{Luca Fabbri\footnote{fabbri@dime.unige.it}}
\affiliation{DIME, Sez. Metodi e Modelli Matematici, Universit\`{a} di Genova,
Via all'Opera Pia 15, 16145 Genova, ITALY}
\date{\today}
\begin{abstract}
We present the theory of Dirac spinors in the formulation given by Bohm on the idea of de Broglie: the quantum relativistic matter field is equivalently re-written as a special type of classical fluid and in this formulation it is shown how a relativistic environment can host the non-local aspects of the above-mentioned hidden-variables theory. Sketches for extensions are given at last.
\end{abstract}
\maketitle
\section{Introduction}
More than one century passed from its beginnings, and yet quantum mechanics still has conceptual issues which have to be addressed at a somewhat fundamental level.

One of the most important is whether wave functions are real or not. A kick in this direction has been recently given in the form of the so-called PBR theorem, pointing out problems that arise from the assumption of the wave function being just information about observables \cite{PBR}.

A real wave function, however, seems to be incompatible with the superposition principle.\! This time hints date back to the EPR argument, suggesting that a wave function should be completed with hidden variables \cite{epr}.

The problem of hidden variables has received a push toward an unexpected direction by Bell, who proved that if hidden variables were indeed a pre-determined feature of the wave function then some very general assumption on the probability distribution would lead to inconsistencies with experiments \cite{B}. Hence, either the probability of a measurement must somehow be influenced by other measurements, or there is no pre-determination. Or both.

One of the first examples of re-formulations of quantum mechanics in which pre-determined hidden variables display the type of non-local behaviour discussed by Bell is the Bohmian version of quantum mechanics \cite{b}. This theory is itself a re-discovery of an older model presented by de Broglie, and so we call it de Broglie-Bohm theory.\footnote{As a matter of fact, this is an instance in which, like in many other cases, the chronological order does not follow the logical order: in fact the de Broglie-Bohm theory was set before the results of Bell, with Bell proving his theorem on the guess that the dBB non-locality could be a general feature of quantum mechanics.}

The dBB theory stems from a re-formulation of quantum mechanics in terms of polar fields, that is when the wave function is written as a real module times a unitary phase. In such a re-formulation the Schr\"{o}dinger equation is split into a Hamilton-Jacobi equation for the ensemble of trajectories and a continuity equation that suggests how we should interpret the velocity of particles. Then a condition of quantum equilibrium is assumed to recover the results of quantum mechanics at a statistical level.

In the dBB theory due to Bohm, the pre-determined hidden variables are the initial positions of particles, and the non-local behaviour can be seen in the fact that the motion of a particle is guided by the universal wave function itself determined by all other particles. When general relativistic constraints are considered, it is not difficult to see where a possible conflict might then arise.

A conflict of this type might be resolved by recovering Lorentz covariance through a foliation of space-time \cite{md}, which may be unobservable and thus not in conflict with relativity in any observation \cite{dgnsz}. Nevertheless, a preferred foliation introduces a privileged time and so it is still incompatible with relativity even if not at practical levels.

A conflict such as the above could also not appear in the first place if we worked in a relativistic version of the dBB theory from the start. That is, instead of asking how the dBB theory can be made relativistic, we ask how an already relativistic theory can be written in dBB form.

We have briefly recalled that the first step to take in order to write the dBB formulation is the polar decomposition of the wave function. And the relativistic theory of matter fields is the well known spinor theory. Therefore, the main aim in this direction should be to consider the Dirac theory and re-write it in polar form.\! This was done in \cite{T1, T2}, and systematically by Takabayasi in \cite{tr1, tr2}.

The application to the dBB theory in the Schr\"{o}dinger case was done in \cite{t} and commented in \cite{bt}. Extensions to include spin and relativistic invariance followed in few works \cite{LdB,Bohm-Schiller,Holland}. However, for spinors, studying relativistic cases does not simply mean allowing velocities to be close to their limit, as we will see. For the Dirac field, the polar form must be investigated more in detail then ever before.

In this paper we bring such a deeper analysis, preparing the Dirac theory to be written in dBB formulation.
\section{The dBB Interpretation}
\subsection{The General Theory}
\subsubsection{The Dirac Fields}
In order to maintain the treatment the most general, we will aim at working in a manifestly covariant environment and in general space-times. This means that in particular electrodynamics, gravity and torsion can also be included if one has the purpose to do so \cite{G}.

To recall the general features of the Dirac theory, we start by assigning the Clifford matrices $\boldsymbol{\gamma}^{a}$ verifying
\begin{eqnarray}
&\{\boldsymbol{\gamma}^{a},\boldsymbol{\gamma}^{b}\}\!=\!2\mathbb{I}\eta^{ab}
\end{eqnarray}
where $\mathbb{I}$ is the identity matrix and
\begin{eqnarray}
&\frac{1}{4}[\boldsymbol{\gamma}^{a},\boldsymbol{\gamma}^{b}]\!=\!\boldsymbol{\sigma}^{ab}
\label{sigma}
\end{eqnarray}
defining $\boldsymbol{\sigma}^{ab}$ from which the matrix $\boldsymbol{\pi}$ is defined as
\begin{eqnarray}
&2i\boldsymbol{\sigma}_{ab}\!=\!\varepsilon_{abcd}\boldsymbol{\pi}\boldsymbol{\sigma}^{cd}
\end{eqnarray}
to set our convention.\footnote{This is usually denoted as a gamma with index five, but it has no sense in the space-time and so we use a notation with no index.}

By exponentiation of the $\boldsymbol{\sigma}^{ab}$ we obtain $\boldsymbol{\Lambda}$ so that
\begin{eqnarray}
&\boldsymbol{S}\!=\!\boldsymbol{\Lambda}e^{iq\alpha}
\label{L}
\end{eqnarray}
is the most complete spinor transformation possible.

Any column of $4$ scalars transforming as
\begin{eqnarray}
&\psi\!\rightarrow\!\boldsymbol{S}\psi
\end{eqnarray}
is a spinor and a row of $4$ scalars transforming as
\begin{eqnarray}
&\overline{\psi}\!\rightarrow\!\overline{\psi}\boldsymbol{S}^{-1}
\end{eqnarray}
is an adjoint spinor. The two are related by
\begin{eqnarray}
&\overline{\psi}\!=\!\psi^{\dagger}\boldsymbol{\gamma}^{0}
\end{eqnarray}
and with them we define the bi-linear quantities as
\begin{eqnarray}
&S^{a}\!=\!\overline{\psi}\boldsymbol{\gamma}^{a}\boldsymbol{\pi}\psi\\
&U^{a}\!=\!\overline{\psi}\boldsymbol{\gamma}^{a}\psi\\
&\Theta\!=\!i\overline{\psi}\boldsymbol{\pi}\psi\\
&\Phi\!=\!\overline{\psi}\psi
\end{eqnarray}
which are all real tensors and such that
\begin{eqnarray}
&U_{a}S^{a}\!=\!0\label{orthogonal1}\\
&U_{a}U^{a}\!=\!-S_{a}S^{a}\!=\!\Theta^{2}\!+\!\Phi^{2}\label{norm1}
\end{eqnarray}
as it is straightforward to demonstrate.

Since the spinorial transformation is point-dependent, we should expect a spinorial gauge connection to emerge.

Indeed one can define the spinorial covariant derivative
\begin{eqnarray}
&\boldsymbol{\nabla}_{\mu}\psi\!=\!\partial_{\mu}\psi\!+\!\frac{1}{2}\Omega_{ij\mu}\boldsymbol{\sigma}^{ij}\psi
\!+\!iqA_{\mu}\psi
\end{eqnarray}
in terms of spin connection and gauge potential in the most general case that respects metric compatibility.

The commutator of spinorial covariant derivatives decomposes according to
\begin{eqnarray}
&[\boldsymbol{\nabla}_{\mu},\boldsymbol{\nabla}_{\nu}]\psi
\!=\!\frac{1}{2}R_{ij\mu\nu}\boldsymbol{\sigma}^{ij}\psi\!+\!iqF_{\mu\nu}\psi
\end{eqnarray}
in terms of the Riemann curvature and Maxwell strength.

Finally, in order to set the dynamical behaviour we are going to consider it to be determined by the Dirac spinor field equations as usual given by 
\begin{eqnarray}
&i\boldsymbol{\gamma}^{\mu}\boldsymbol{\nabla}_{\mu}\psi
\!-\!XW_{\mu}\boldsymbol{\gamma}^{\mu}\boldsymbol{\pi}\psi\!-\!m\psi\!=\!0
\label{D}
\end{eqnarray}
in which $W_{\mu}$ is the axial-vector Hodge dual of the torsion tensor and $X$ the torsion-spin coupling, which has been added to be in the most general situation possible.

General definitions can be taken for instance from \cite{G}.
\subsubsection{Full Geometric Coupling}
So far we have given the Dirac theory in full coupling, that is when the Dirac equation is written in presence of electrodynamics, gravity and torsion. Hence to complete the theory, we now give also the geometric equations determining electrodynamics, gravity and torsion sourced by the Dirac field. This will also be useful later on, when we will begin the study of classical approximation.

So, given the Dirac spinor matter field, it sources electrodynamics in terms of the Maxwell equations
\begin{eqnarray}
&\nabla_{\sigma}F^{\sigma\mu}\!=\!qU^{\mu}
\label{ee}
\end{eqnarray}
where the right-hand side can be written with the velocity and $\nabla_{\alpha}U^{\alpha}\!=\!0$ due to the validity of the Dirac equations.

The source to torsion dynamics is instead
\begin{eqnarray}
&\nabla_{\sigma}(\partial W)^{\sigma\mu}\!+\!M^{2}W^{\mu}\!=\!XS^{\mu}
\label{te}
\end{eqnarray}
with $(\partial W)_{\sigma\mu}\!=\!\partial_{[\sigma}W_{\mu]}$ and $M$ the mass of torsion and in which the right-hand side can be written with the spin so $M^{2}\nabla_{\mu}W^{\mu}\!=\!2X\Theta m$ due to the Dirac equations.

The source of gravity is described by
\begin{eqnarray}
&R^{\rho\sigma}\!-\!\frac{1}{2}Rg^{\rho\sigma}
\!-\!\Lambda g^{\rho\sigma}\!=\!\frac{1}{2}T^{\rho\sigma}
\label{ge}
\end{eqnarray}
in which $R^{\alpha\rho\mu\sigma}g_{\alpha\mu}\!=\!R^{\rho\sigma}$ and $R^{\rho\sigma}g_{\rho\sigma}\!=\!R$ are the Ricci tensors and $\Lambda$ the cosmological constant and in which the right-hand side is given according to the energy
\begin{eqnarray}
\nonumber
&T^{\rho\sigma}\!=\!\frac{1}{4}F^{2}g^{\rho\sigma}
\!-\!F^{\rho\alpha}\!F^{\sigma}_{\phantom{\sigma}\alpha}+\\
\nonumber
&+\frac{1}{4}(\partial W)^{2}g^{\rho\sigma}
\!-\!(\partial W)^{\sigma\alpha}(\partial W)^{\rho}_{\phantom{\rho}\alpha}+\\
\nonumber
&+M^{2}(W^{\rho}W^{\sigma}\!-\!\frac{1}{2}W^{2}g^{\rho\sigma})+\\
\nonumber
&+\frac{i}{4}(\overline{\psi}\boldsymbol{\gamma}^{\rho}\boldsymbol{\nabla}^{\sigma}\psi
\!-\!\boldsymbol{\nabla}^{\sigma}\overline{\psi}\boldsymbol{\gamma}^{\rho}\psi+\\
\nonumber
&+\overline{\psi}\boldsymbol{\gamma}^{\sigma}\boldsymbol{\nabla}^{\rho}\psi
\!-\!\boldsymbol{\nabla}^{\rho}\overline{\psi}\boldsymbol{\gamma}^{\sigma}\psi)-\\
&-\frac{1}{2}X(W^{\sigma}\overline{\psi}\boldsymbol{\gamma}^{\rho}\boldsymbol{\pi}\psi
\!+\!W^{\rho}\overline{\psi}\boldsymbol{\gamma}^{\sigma}\boldsymbol{\pi}\psi)
\label{energy}
\end{eqnarray}
so $\nabla_{\alpha}T^{\alpha\nu}\!=\!0$ and with trace as $2R\!+\!8\Lambda\!-\!M^{2}W^{2}\!=\!-\Phi m$ both verified due to the validity of the Dirac equations.
\subsubsection{The Polar Form}
Having recalled the general definitions of the Dirac theory, we next convert it into its polar form \cite{tr1,tr2}. Just the same, we will follow a different route as compared to Takabayasi. Our goal is writing the polar form of a theory displaying not only relativistic covariance, but also manifest covariance under general curvilinear coordinates in curved space-times, as well as gauge covariance under the local Lorentz\! transformations.\! Readers interested in more details can find them in \cite{Fabbri:2020ypd} and references therein.

The main idea that lies behind the polar decomposition is that each component of the spinor be re-written as the product of a module times a unitary phase. Because all components mix during a spinor transformation, such a decomposition in general does not respect manifest covariance, unless due care is taken. When this is done, it is possible to find that the most general spinor field can always be written in chiral representation as
\begin{eqnarray}
&\!\psi\!=\!\phi e^{-\frac{i}{2}\beta\boldsymbol{\pi}}
\boldsymbol{L}^{-1}\left(\begin{tabular}{c}
$1$\\
$0$\\
$1$\\
$0$
\end{tabular}\right)
\label{spinor}
\end{eqnarray}
with $\phi$ and $\beta$ being a real scalar and a real pseudo-scalar fields known as module and chiral angle, and where $\boldsymbol{L}$ is a general spinor transformation. In this form we have
\begin{eqnarray}
&\Theta\!=\!2\phi^{2}\sin{\beta}\\
&\Phi\!=\!2\phi^{2}\cos{\beta}
\end{eqnarray}
with
\begin{eqnarray}
&S^{a}\!=\!2\phi^{2}s^{a}\\
&U^{a}\!=\!2\phi^{2}u^{a}
\end{eqnarray}
such that
\begin{eqnarray}
&u_{a}s^{a}\!=\!0\\
&u_{a}u^{a}\!=\!-s_{a}s^{a}\!=\!1
\end{eqnarray}
are the velocity vector and spin axial-vector. This shows that module and chiral angle are the only true degrees of freedom whereas the spatial components of velocity and spin can always be boosted to zero or rotated to point along a given direction. In polar form the components of the spinor are re-arranged so that $\phi$ and $\beta$ are isolated from the parameters of the spinor transformation $\boldsymbol{L}$ and because these can always be transferred into the frame they can be recognized to be the Goldstone fields.\footnote{Notice that such a polar decomposition is always possible so long as $\Theta$ and $\Phi$ are not identically zero as it generally happens. In the specific circumstance in which $\Theta^{2}\!+\!\Phi^{2}\!\equiv\!0$ we would still have a polar decomposition. However, in this case the fields would be pure Goldstone states. Therefore, we are not going to consider this singular case in the following of this paper.}

To the best of our knowledge, the first appearance of the polar form (\ref{spinor}) in literature was in \cite{jl}. When this polar form is written with (\ref{L}) we have the more explicit expression given with all fields as
\begin{eqnarray}
&\!\psi\!=\!\phi\ e^{-iq\alpha} e^{-\frac{i}{2}\beta\boldsymbol{\pi}}
\boldsymbol{\Lambda}^{-1}\left(\begin{tabular}{c}
$1$\\
$0$\\
$1$\\
$0$
\end{tabular}\right)
\end{eqnarray}
in which $\phi e^{-iq\alpha}$ is the single global factor we would have had in the usual case but now we also have a chiral phase $e^{-i\beta\boldsymbol{\pi}/2}$ acting on the two chiral projections in opposite ways plus the complex Lorentz transformation $\boldsymbol{\Lambda}^{-1}$ accounting for boosts and rotations mixing each helicity in an independent manner. These two last elements are not usually addressed in studying spinors, and this might be the missing piece in the analysis of Takabayasi \cite{tr1,tr2}.

Let us now move on to study the differential structures in polar form. With a little algebra that we are not going to reproduce, one can show that we can always write
\begin{eqnarray}
\boldsymbol{L}^{-1}\partial_{\mu}\boldsymbol{L}\!=\!iq\partial_{\mu}\xi\mathbb{I}
\!+\!\frac{1}{2}\partial_{\mu}\xi^{ab}\boldsymbol{\sigma}_{ab}\label{LdL}
\end{eqnarray}
for some $\xi$ and $\xi^{ab}$ which are indeed the Goldstone states of the spinor field.\! Using this expression in the polar form of the spinorial covariant derivative, one can set
\begin{eqnarray}
&\partial_{\mu}\xi_{ij}\!-\!\Omega_{ij\mu}\!\equiv\!R_{ij\mu}\label{R}\\
&q(\partial_{\mu}\xi\!-\!A_{\mu})\!\equiv\!P_{\mu}\label{P}
\end{eqnarray}
in terms of which
\begin{eqnarray}
&\!\!\!\!\!\!\!\!\boldsymbol{\nabla}_{\mu}\psi\!=\!(-\frac{i}{2}\nabla_{\mu}\beta\boldsymbol{\pi}
\!+\!\nabla_{\mu}\ln{\phi}\mathbb{I}
\!-\!iP_{\mu}\mathbb{I}\!-\!\frac{1}{2}R_{ij\mu}\boldsymbol{\sigma}^{ij})\psi
\label{decspinder}
\end{eqnarray}
for the spinor field in the most general case. From this
\begin{eqnarray}
&\nabla_{\mu}s_{i}\!=\!R_{ji\mu}s^{j}\label{ds}\\
&\nabla_{\mu}u_{i}\!=\!R_{ji\mu}u^{j}\label{du}
\end{eqnarray}
in general. Before we have seen that it is always possible to have the Goldstone states transferred into gauge and frames, and now we can see what happens to them. They are absorbed by spin connection and gauge potential as the longitudinal components of $P_{\mu}$ and $R_{ji\mu}$ which are a real vector and a tensor respectively, so that they have the same information of gauge potential and spin connection but they are gauge invariant and frame covariant.

So long as we can be aware, Jakobi and Lochak did not continue the investigation of the polar form at a differential level while still maintaining manifest covariance, so that we believe that the definitions of tensors (\ref{R}-\ref{P}), the proof of their frame and gauge covariance, as well as the form of the derivative (\ref{decspinder}), are all new results \cite{Fabbri:2018crr}.

The Riemann curvature and Maxwell strength are then
\begin{eqnarray}
&\!\!\!\!\!\!\!\!R^{i}_{\phantom{i}j\mu\nu}\!=\!-(\nabla_{\mu}R^{i}_{\phantom{i}j\nu}
\!-\!\!\nabla_{\nu}R^{i}_{\phantom{i}j\mu}
\!\!+\!R^{i}_{\phantom{i}k\mu}R^{k}_{\phantom{k}j\nu}
\!-\!R^{i}_{\phantom{i}k\nu}R^{k}_{\phantom{k}j\mu})\label{Riemann}\\
&\!\!\!\!qF_{\mu\nu}\!=\!-(\nabla_{\mu}P_{\nu}\!-\!\nabla_{\nu}P_{\mu})\label{Faraday}
\end{eqnarray}
identically. Above we have remarked that in $P_{\mu}$ and $R_{ji\mu}$ we find the same information of gauge potential and spin connection although these are gauge invariant and frame covariant. As Riemann curvature and Maxwell strength contain information about gravity and electrodynamics only, then non-zero $R_{ji\mu}$ and $P_{\mu}$ that are solutions of the $R^{i}_{\phantom{i}j\mu\nu}\!=\!0$ and $F_{\mu\nu}\!=\!0$ conditions encode the information about spin connection and gauge potential that is not gravitational and electrodynamic, respectively.

That $R^{i}_{\phantom{i}j\mu\nu}\!=\!0$ and $F_{\mu\nu}\!=\!0$ can have non-trivial solutions was proven with a direct example in \cite{Fabbri:2019kfr}.

Introducing the combined potential and its Hodge dual
\begin{eqnarray}
&\Sigma_{ij\mu}\!=\!R_{ij\mu}\!-\!2P_{\mu}u^{a}s^{b}\varepsilon_{ijab}
\label{Sigma}\\
&M^{ab}_{\phantom{ab}\mu}\!=\!\frac{1}{2}R_{ij\mu}\varepsilon^{ijab}\!+\!2P_{\mu}u^{[a}s^{b]}
\label{M}
\end{eqnarray}
it is possible to see that the Dirac equations in polar form are equivalently written according to
\begin{eqnarray}
&\nabla_{\mu}\beta\!-\!2XW_{\mu}\!+\!M_{\mu}\!+\!2ms_{\mu}\cos{\beta}\!=\!0\label{dep1}\\
&\nabla_{\mu}\ln{\phi^{2}}\!+\!\Sigma_{\mu}\!+\!2ms_{\mu}\sin{\beta}\!=\!0\label{dep2}
\end{eqnarray}
specifying all the derivatives of module and chiral angle.

Once again, to the best of our knowledge, the very first appearance of the polar form of the Dirac equations was in \cite{y}, although we are not aware of any work of Yvon or subsequent authors in which the polar form of the Dirac equations was written in a manifestly covariant way and for whatever potential, as it is done in equations (\ref{dep1}, \ref{dep2}).

The interested readers can find more details in \cite{Fabbri:2020ypd}.
\subsubsection{Relativistic Quantum Potentials}
So far we wrote the spinorial field in polar form. We are now going to assign to its most fundamental elements the corresponding interpretation that we would have in the relativistic version of the dBB theory. Being in the case of complete generality, some element will remain obscure, and so we ask the reader some patience. The immediately following section will treat the non-relativistic limit, and there the full correspondence will become obvious.

So, to summarize the results obtained in the previous sub-section, in the scheme of the dBB interpretation, we can say that, once the Dirac equations are assigned, and the polar form used, we can write them, after defining a pair of dual potentials (\ref{Sigma}-\ref{M}), according to (\ref{dep1}-\ref{dep2}). Yet, we could alternatively define also the following potentials
\begin{eqnarray}
&2Y_{\mu}\!=\!\nabla_{\mu}\beta\!-\!2XW_{\mu}
\!+\!\frac{1}{2}\varepsilon_{\mu\nu\alpha\sigma}R^{\nu\alpha\sigma}
\label{Y}\\
&-2Z_{\mu}\!=\!\nabla_{\mu}\ln{\phi^{2}}\!+\!R_{\mu\nu}^{\phantom{\mu\nu}\nu}
\label{Z}
\end{eqnarray}
in which $P^{\mu}$ has been left out so that
\begin{eqnarray}
&P^{\iota}u_{[\iota}s_{\mu]}\!-\!Y_{\mu}
\!-\!ms_{\mu}\cos{\beta}\!=\!0\label{ddb1}\\
&P^{\rho}u^{\nu}s^{\alpha}\varepsilon_{\mu\rho\nu\alpha}\!+\!Z_{\mu}
\!-\!ms_{\mu}\sin{\beta}\!=\!0\label{ddb2}
\end{eqnarray}
and these have to be recognized as the Hamilton-Jacobi equations of the dBB interpretation. In fact, (\ref{Y}-\ref{Z}) are objects containing the derivatives of the degrees of freedom, and as such they are the quantum potentials in the relativistic case with spin (they contain only one derivative as clear from relativistic covariance and they are two because of the presence of two chiral fields). They do not contain the object $P_{\mu}$ which can then be found only in (\ref{ddb1}-\ref{ddb2}), and if $P_{\mu}$ could be identified with the momentum then (\ref{ddb1}-\ref{ddb2}) would be equations yielding the structure of the momentum, in terms of the quantum potentials, and as a consequence of this fact they would result to be the Hamilton-Jacobi equations, by their very definition.

To see whether $P_{\mu}$ can be identified with the momentum, we combine (\ref{ddb1}-\ref{ddb2}) and manipulate them so to get
\begin{eqnarray}
&\!\!\!\!P^{\rho}\!=\!mu^{\rho}\cos{\beta}
\!+\!\left(Y^{\iota}u_{\iota}g^{\rho\alpha}\!-\!Y^{\alpha}u^{\rho}
\!+\!Z_{\mu}u_{\nu}\varepsilon^{\mu\nu\rho\alpha}\right)\!s_{\alpha}
\label{momentum}
\end{eqnarray}
which gives its explicit expression. We see that $P_{\mu}$ equals the kinematic momentum $mu_{\mu}$ up to a multiplicative factor of the chiral angle $\cos{\beta}$ and plus corrections proportional the product of the spin and the potentials.

It is important to remark that here the interpretation of the relativistic probability amplitude as given by $2\phi^{2}$ seems to arise quite naturally. It clearly could neither be the temporal component of $U^{a}$ (which is not covariant) nor $\Phi$ (which is not positive defined). However, because in relativistic spinning cases we have a second scalar\! $\Theta$ it is possible to write the scalar $\sqrt{\Phi^{2}\!+\!\Theta^{2}}$ which is positive defined. But this is simply $2\phi^{2}$ as we argued above.

Notice that the above quantum potentials are a relativistic spinning version of what Bohm would call quantum potentials. The fact that they are relativistic is evident in their being first-order differential (because Dirac is at first-order derivative while Schr\"{o}dinger is at second-order derivative) and that they describe spinning systems is clear from the existence of two of them (since solutions of the Dirac equations are characterized by two degrees of freedom whereas solutions of the Schr\"{o}dinger equation are characterized by one degree of freedom).
\subsection{Two Special Limiting Cases}
\subsubsection{Non-Relativistic Chiral Limit}
The polar decomposition of spinors is very much linked to the dimension of the space that contains them. So the polar form (\ref{spinor}) can only be valid in the $(1\!+\!3)$-dimensional space-times. In other dimensions or signatures we would have a different polar decomposition \cite{Fabbri:2019nks}. In particular, for the $3$-dimensional space we would have a polar form given according to the following structure
\begin{eqnarray}
&\!\psi\!=\!\phi \boldsymbol{R}^{-1}\left(\begin{tabular}{c}
$1$\\
$0$
\end{tabular}\right)
\label{spinorino}
\end{eqnarray}
where $\phi$ is the module, and $\boldsymbol{R}$ is a general complex rotation. In this form we would have that
\begin{eqnarray}
&\Phi\!=\!\psi^{\dagger}\psi\!=\!\phi^{2}
\end{eqnarray}
with
\begin{eqnarray}
&\vec{S}\!=\!\psi^{\dagger}\vec{\boldsymbol{\sigma}}\psi\!=\!\phi^{2}\vec{s}
\end{eqnarray}
such that
\begin{eqnarray}
&\vec{s}\!\cdot\!\vec{s}\!=\!1
\end{eqnarray}
as a constraint on the spin vector.\footnote{We assume the reader familiar with the Pauli matrices $\vec{\boldsymbol{\sigma}}$ above.}

The full expression in presence of a phase is given by
\begin{eqnarray}
&\!\psi\!=\!\phi e^{-iq\alpha}\boldsymbol{R}^{-1}\left(\!\begin{tabular}{c}
$1$\\
$0$
\end{tabular}\!\right)
\end{eqnarray}
but we notice that even now we cannot accommodate the unitary chiral phase as well as the boosts.

The non-relativistic limit is implemented by requiring that boosts be disallowed and that time be excluded as a dimension, so that we are essentially asking that spinors in $(1\!+\!3)$-dimensional space-times be reduced to spinors in $3$-dimensional spaces. Hence (\ref{spinor}) must somehow reduce to (\ref{spinorino}).\! Writing\! (\ref{spinor}) in standard representation makes us see that it does reduce to (\ref{spinorino}) whenever we have
\begin{eqnarray}
&\beta\!\rightarrow\!0\label{nrb}\\
&\vec{u}\!\rightarrow\!0\label{nru}
\end{eqnarray}
as is discussed in \cite{Fabbri:2020ypd}. The passage from relativistic cases to non-relativistic cases is not only $\vec{u}\!\rightarrow\!0$ when considering spinors because for them also $\beta\!\rightarrow\!0$ has to be imposed for consistency. Lack of doing so will not ensure, even in the rest frame, the non-relativistic limit. Henceforth, we may think at $\beta$ as what contains the information on the internal dynamics of spinors. Notice however that these two conditions (\ref{nrb}-\ref{nru}) are together equivalent to asking that when written in standard representation the spinor lose its small components \cite{Fabbri:2020ypd}. This last is precisely the definition of non-relativistic limit that is commonly used in mathematical physics. Notice that the non-relativistic limit has been treated without involving the momentum.
\subsubsection{Non-Quantum Helicity Limit}
Let us now consider the expression of the energy (\ref{energy}) written in polar form, and with the momentum (\ref{momentum}) substituted in it, which then becomes
\begin{eqnarray}
\nonumber
&T^{\rho\sigma}\!=\!\frac{1}{4}F^{2}g^{\rho\sigma}
\!-\!F^{\rho\alpha}\!F^{\sigma}_{\phantom{\sigma}\alpha}+\\
\nonumber
&+\frac{1}{4}(\partial W)^{2}g^{\rho\sigma}
\!-\!(\partial W)^{\sigma\alpha}(\partial W)^{\rho}_{\phantom{\rho}\alpha}+\\
\nonumber
&+M^{2}(W^{\rho}W^{\sigma}\!-\!\frac{1}{2}W^{2}g^{\rho\sigma})+\\
&+2\phi^{2}m\cos{\beta}u^{\sigma}u^{\rho}\!+\!E^{\rho\sigma\kappa}s_{\kappa}
\end{eqnarray}
where we have introduced
\begin{eqnarray}
\nonumber
&E^{\rho\sigma\kappa}\!=\!\phi^{2}[g^{\rho\kappa}Y^{\sigma}\!+\!g^{\sigma\kappa}Y^{\rho}
\!-\!2Y^{\kappa}u^{\sigma}u^{\rho}+\\
\nonumber
&+Y^{\iota}u_{\iota}u^{\rho}g^{\sigma\kappa}\!+\!Y^{\iota}u_{\iota}u^{\sigma}g^{\rho\kappa}+\\
\nonumber
&+Z_{\mu}u_{\nu}u^{\rho}\varepsilon^{\mu\nu\sigma\kappa}
\!+\!Z_{\mu}u_{\nu}u^{\sigma}\varepsilon^{\mu\nu\rho\kappa}-\\
\nonumber
&-\frac{1}{4}(R_{\alpha\nu}^{\phantom{\alpha\nu}\sigma}\varepsilon^{\rho\alpha\nu\kappa}
\!+\!R_{\alpha\nu}^{\phantom{\alpha\nu}\rho}\varepsilon^{\sigma\alpha\nu\kappa}+\\
&+\varepsilon^{\rho\nu\alpha\iota}g^{\sigma\kappa}R_{\nu\alpha\iota}
\!+\!\varepsilon^{\sigma\nu\alpha\iota}g^{\rho\kappa}R_{\nu\alpha\iota})]
\end{eqnarray}
for compactness. By taking its divergence we have
\begin{eqnarray}
\nonumber
&2\phi^{2}u^{\rho}\nabla_{\rho}(m\cos{\beta}u^{\sigma})
\!+\!(\nabla_{\rho}E^{\rho\sigma\alpha}
\!+\!E^{\rho\sigma\kappa}R^{\alpha}_{\phantom{\alpha}\kappa\rho})s_{\alpha}=\\
&=2\phi^{2}[qF^{\sigma\alpha}u_{\alpha}\!+\!X(\partial W)^{\sigma\alpha}s_{\alpha}
\!-\!2Xm\sin{\beta}W^{\sigma}]
\label{New}
\end{eqnarray}
and this will be recognized as the Newton law.

To see this, we consider that the non-quantum limit is implemented by the condition of spinlessness
\begin{eqnarray}
&s^{i}\!\rightarrow\!0\label{nq}
\end{eqnarray}
and as $\nabla S\!=\!2m\Theta$ we have $\beta\!\rightarrow\!0$ in the same limit.\footnote{The non-quantum limit would be implemented by $\hbar\!\rightarrow\!0$ which is hidden in our presentation with natural units. If we were not to assume them, $\hbar\!\rightarrow\!0$ would clearly give the spinless condition.}

In this classical approximation (\ref{New}) reduces to 
\begin{eqnarray}
&2\phi^{2}u^{\rho}\nabla_{\rho}P^{\sigma}\!=\!2\phi^{2}qF^{\sigma\alpha}u_{\alpha}
\end{eqnarray}
since $P^{\mu}\!=\!mu^{\mu}$ in the same regime.

Simplifying the module would give
\begin{eqnarray}
&u^{\eta}\nabla_{\eta}P^{\sigma}\!=\!qF^{\sigma\alpha}u_{\alpha}
\label{Newton}
\end{eqnarray}
which can finally be seen as the Newton law of motion.

In its derivation, we never assumed constraints on the matter distribution. In spinless situations, all points follow the classical motion, not only the peak of a localized matter distribution.\! We regard this as an improvement in comparison with Ehrenfest theorem, where the material distribution is localized and only its peak follows classical trajectories. We now move to the dBB interpretation.
\subsection{Recovery of the dBB Model}
\subsubsection{Second-Order Differential Field Equations}
Let us next consider the Dirac equations in polar form as given by (\ref{dep1}-\ref{dep2}),\! and let us apply them onto each other, so to eliminate the presence of velocity and spin.

The second-order differential equations are hence
\begin{eqnarray}
\nonumber
&\nabla^{\mu}(\phi^{2}\nabla_{\mu}\beta)
\!-\!(8X^{2}M^{-2}\phi^{2}m\sin{\beta}-\\
&-2XW^{\nu}\Sigma_{\nu}\!-\!\nabla_{\nu}M^{\nu}
\!+\!M^{\nu}\Sigma_{\nu})\phi^{2}\!=\!0
\end{eqnarray}
and
\begin{eqnarray}
\nonumber
&\left|\nabla\beta/2\right|^{2}\!\!-\!m^{2}\!-\!\phi^{-1}\nabla^{2}\phi
\!+\!\frac{1}{4}(-2\nabla_{\nu}\Sigma^{\nu}+\\
&+\Sigma^{\nu}\Sigma_{\nu}\!-\!M^{\nu}M_{\nu}
\!+\!4XW_{\nu}M^{\nu}\!-\!4X^{2}W_{\nu}W^{\nu})\!=\!0\label{General}
\end{eqnarray}
in terms of the external potentials of torsion, gravity and electrodynamics.\! In the following we will work in the case of the $\beta\!\rightarrow\!0$ limit and focus on the second of them.

As a start, we see that when torsion is in its effective approximation, the torsion field equations reduce to
\begin{eqnarray}
&M^{2}W^{\mu}\!\approx\!XS^{\mu}
\end{eqnarray}
which can then be substituted into the above (\ref{General}) to get
\begin{eqnarray}
\nonumber
&\nabla^{2}\phi\!-\!4X^{4}M^{-4}\phi^{5}
\!-\!2X^{2}M^{-2}M^{\nu}s_{\nu}\phi^{3}+\\
&+\frac{1}{4}(2\nabla_{\nu}\Sigma^{\nu}\!-\!\Sigma^{\nu}\Sigma_{\nu}
\!+\!M^{\nu}M_{\nu}\!+\!4m^{2})\phi\!=\!0\label{Effective}
\end{eqnarray}
which is remarkably non-linear. Notice that if $M^{\nu}s_{\nu}\!>\!0$ the effective self-interaction remains attractive at smaller densities and if $\nabla_{\nu}\Sigma^{\nu}/2\!-\!\Sigma^{\nu}\Sigma_{\nu}/4\!+\!M^{\nu}M_{\nu}/4\!+\!m^{2}\!>\!0$ the effective mass term remains positive. If both conditions are verified the above equation acquires a structure that can allow solitonic solutions. It is tempting to suggest an interpretation for which this localized distribution could represent the particle, instead of postulating it \emph{ad hoc}.

Far from the peak, the non-linear terms tend to vanish rapidly, so more simplifications occur. For the additional assumption $R_{ij\nu}\!\approx\!0$ (\ref{Effective}) can be written like
\begin{eqnarray}
&P^{2}\!-\!m^{2}\!-\!\frac{1}{2}qF_{\mu\nu}u_{\rho}s_{\sigma}\varepsilon^{\mu\nu\rho\sigma}
\!-\!\phi^{-1}\nabla^{2}\phi\!=\!0\label{standard}
\end{eqnarray}
which is the equation that constitutes the balance of energy with an external potential given in terms of the electrodynamic coupling plus the quantum potential $\phi^{-1}\nabla^{2}\phi$ in the form known from the standard dBB interpretation.

In non-relativistic limit and setting $P^{0}\!-\!m\!=\!H$ we have
\begin{eqnarray}
H\!=\!\frac{1}{2m}\vec{P}\!\cdot\!\vec{P}
\!-\!\frac{q}{m}\frac{\vec{s}}{2}\!\cdot\!\vec{B}
\!-\!\frac{1}{2m}\phi^{-1}\vec{\nabla}\!\cdot\!\vec{\nabla}\phi
\end{eqnarray}
as the Hamiltonian with $F_{IJ}\!=\!-\varepsilon_{IJK}B^{K}$ and giving and quantum potentials precisely as in the dBB formulation.

The non-relativistic limit is gotten from the guidance equation (\ref{momentum}) which under the above hypotheses is
\begin{eqnarray}
&P^{\rho}\!=\!mu^{\rho}
\!-\!\frac{1}{2}\varepsilon^{\rho\nu\alpha\mu}u_{\nu}s_{\alpha}\nabla_{\mu}\ln{\phi^{2}}
\end{eqnarray}
therefore giving that $P^{0}\!\rightarrow\!m$ as is discussed above as well as $\vec{P}\!=\!\vec{s}\!\times\!\vec{\nabla}\ln{\phi}$ if we neglect the time dependence of the module. Notice that for a module of gaussian distribution $\phi\!=\!K\exp{(-kr^{2}/8)}$ we have $\vec{\nabla}\!\times\!\vec{P}\!=\!k\vec{s}/2$ showing that a matter distribution of this type converges only if the curl of its momentum is directed along its spin axial-vector.

Notice finally that in the same limit we also have that $2\phi^{2}\!\equiv\!\sqrt{\Phi^{2}\!+\!\Theta^{2}}\!\rightarrow\!|\Phi|$ which is the probability amplitude of non-relativistic quantum mechanics, as known.
\section{Free Will}
\subsection{Physical Contextuality}
As we had the opportunity to mention above, the dBB model is one of the first in which an explicitly non-local behaviour was found. And in fact, it was the prototypical model that had led Bell to ask whether this property was a general feature of quantum mechanics. In detail, Bell's argument relies on the definition of a form of non-locality, known as Bell non-locality, which can be used to deduce specific constraints, called Bell inequalities \cite{B}. All along the years many forms of Bell-like inequalities have been proposed. A very general one is what is known as CHSH inequality \cite{CHSH}. Generally, the specific type of inequality is irrelevant, as they all have in common the idea that a Bell inequality is the manifestation of the pre-determined hidden variables. If a theory has pre-determined hidden variables then, by its structure, it must imply some form of pattern in measurements, and this pattern is reflected as inequalities between the results of observations.

Because quantum mechanics does not verify these inequalities, then either there are no pre-determined hidden variables or if they do exist then they cannot be local.

In time, various generalizations of this statement have been proposed. The first one is due to Bell himself \cite{Bell}.\footnote{In fact, \cite{Bell} even pre-dates \cite{B}.}

As compared to \cite{B}, in \cite{Bell} the accent is shifted, from the concept of non-locality, to that of contextuality. That is, the fact that the wave function of one particle has to include variables pertaining to other particles, even with space-like separation, is seen as a more general statement about the fact that the result of a measurement depends on which other measurement is chosen to be made within the settings of the experimental apparatus \cite{KS,FW}.

The general statement however is similar, and so either there are no pre-determined hidden variables or the result of a measurement must depend on other measurements.

Frequently \cite{Bell,KS} are seen as two parallel versions of the same theorem, while \cite{FW} has more of an independent formulation in its involving a definition of free will.

Free will, in \cite{FW}, is defined as the lack of determinism in the sense of leisure of choosing the experimental setting at the convenience of observers. In what follows, we wish to provide a less (experimentally) practical though more (theoretically) precise definition of the in-determinism of physical processes. Because some physical effects are determined, it is unwise to talk about in-determinism, and we will consider the more sober term under-determinism.

So a mathematically finer \emph{under-determinism might be defined as the fact that in physics there exist effects that are not dynamically determined as solutions of differential field equations with an external source term}. Notice that this definition is compatible with our idea that when, on the contrary, an effect can be seen as described by the solution of a field equations with a source then there is no freedom of choice once the source is assigned. Remark however that to be more precise this definition does not account for the freedom of choosing boundary conditions, although there is no physical theory that does specify a choice of boundary conditions, so that we will not discuss this circumstance in the following of the presentation.
\subsection{Dynamical Under-Determination}
In the first section we have demonstrated how we can have the spinor field written in polar decomposition, with which it becomes possible to infer the dBB interpretation.

Now we would like to consider the same theory without the assumptions that led to the dBB interpretation, and that is keeping $\beta$, $W_{\mu}$ as well as $R_{ij\nu}$ non-zero. The most important of these objects for our purposes are the $R_{ij\nu}$ tensors, as they are still not well understood. The present section is devoted to study them more in detail and find a link with the definition of under-determination above.

We start by investigating the general structure of the spinor field. As we said above, the polar form (\ref{spinor}) allows to keep the degrees of freedom $\phi$ and $\beta$ isolated from the Goldstone states of the system contained in $\boldsymbol{L}$ and which can therefore be transferred into gauge and frame. This field describing the Goldstone states is actually given by its derivative $\boldsymbol{L}^{-1}\partial_{\mu}\boldsymbol{L}$ as this is the field that will combine with the gauge potential and spin connection to yield the $P_{\nu}$ and $R_{ij\nu}$ tensors. Notice that as Goldstone states, $\boldsymbol{L}$, and therefore $\boldsymbol{L}^{-1}\partial_{\mu}\boldsymbol{L}$, contains information about gauge and frames.\! And it is of course not covariant.\! Because its curvature tensor is given by
\begin{eqnarray}
\nonumber
&\partial_{\mu}(\boldsymbol{L}^{-1}\partial_{\nu}\boldsymbol{L})
\!-\!\partial_{\nu}(\boldsymbol{L}^{-1}\partial_{\mu}\boldsymbol{L})+\\
\nonumber
&+(\boldsymbol{L}^{-1}\partial_{\mu}\boldsymbol{L})
(\boldsymbol{L}^{-1}\partial_{\nu}\boldsymbol{L})
\!-\!(\boldsymbol{L}^{-1}\partial_{\nu}\boldsymbol{L})
(\boldsymbol{L}^{-1}\partial_{\mu}\boldsymbol{L})\equiv\\
\nonumber
&\equiv\partial_{\mu}\boldsymbol{L}^{-1}\partial_{\nu}\boldsymbol{L}
\!+\!\boldsymbol{L}^{-1}\partial_{\mu}\partial_{\nu}\boldsymbol{L}-\\
\nonumber
&-\partial_{\nu}\boldsymbol{L}^{-1}\partial_{\mu}\boldsymbol{L}
\!-\!\boldsymbol{L}^{-1}\partial_{\nu}\partial_{\mu}\boldsymbol{L}-\\
\nonumber
&-\boldsymbol{L}^{-1}\boldsymbol{L}
\partial_{\mu}\boldsymbol{L}^{-1}\partial_{\nu}\boldsymbol{L}
\!+\!\boldsymbol{L}^{-1}\boldsymbol{L}
\partial_{\nu}\boldsymbol{L}^{-1}\partial_{\mu}\boldsymbol{L}=\\
&=\boldsymbol{L}^{-1}\partial_{\mu}\partial_{\nu}\boldsymbol{L}
\!-\!\boldsymbol{L}^{-1}\partial_{\nu}\partial_{\mu}\boldsymbol{L}\!=\!0
\end{eqnarray}
then they contain no information about electrodynamics and gravity. On the other hand, the gauge potentials and spin connection contain information both on gauge and frames and on electrodynamics and\! gravity.\! To split them one considers the curvatures. Thus $A_{\nu}$ such that $F_{\mu\nu}\!=\!0$ contains information about gauge but not electrodynamics whereas\! $\Omega_{ij\nu}$ such\! that\! $R_{ij\mu\nu}\!=\!0$ contains information about frames but not gravity. Once again, $A_{\nu}$ and $\Omega_{ij\nu}$ are non-covariant objects. Notice nevertheless that their information about electrodynamics and gravity is also in the Maxwell strength and Riemann curvature. And these two tensors are of course covariant. When the Goldstone fields combine with spin connection and gauge potentials as in (\ref{R}-\ref{P}) all non-covariant properties cancel off. To see it, consider that for spinor transformations, nothing within the spinor in polar form can transforms except
\begin{eqnarray}
&\boldsymbol{L}^{-1}\!\rightarrow\!\boldsymbol{S}\boldsymbol{L}^{-1}
\end{eqnarray}
as it might have been expected. The transformation law of the spinor connection is
\begin{eqnarray}
\boldsymbol{\Omega}_{\mu}\!\rightarrow\!\boldsymbol{S}\left(\boldsymbol{\Omega}_{\mu}
-\boldsymbol{S}^{-1}\partial_{\mu}\boldsymbol{S}\right)\boldsymbol{S}^{-1}
\end{eqnarray}
as also well known. Consequently
\begin{eqnarray}
\nonumber
&\partial_{\mu}\boldsymbol{L}^{-1}\boldsymbol{L}\!+\!\boldsymbol{\Omega}_{\mu}
\!\rightarrow\!\partial_{\mu}(\boldsymbol{L}\boldsymbol{S}^{-1})^{-1}(\boldsymbol{L}\boldsymbol{S}^{-1})+\\
\nonumber
&+\boldsymbol{S}\left(\boldsymbol{\Omega}_{\mu}
-\boldsymbol{S}^{-1}\partial_{\mu}\boldsymbol{S}\right)\boldsymbol{S}^{-1}
\!=\!\partial_{\mu}\boldsymbol{S}\boldsymbol{L}^{-1}\boldsymbol{L}\boldsymbol{S}^{-1}+\\
\nonumber
&+\boldsymbol{S}\partial_{\mu}\boldsymbol{L}^{-1}\boldsymbol{L}\boldsymbol{S}^{-1}
\!+\!\boldsymbol{S}\boldsymbol{\Omega}_{\mu}\boldsymbol{S}^{-1}
\!-\!\partial_{\mu}\boldsymbol{S}\boldsymbol{S}^{-1}=\\
\nonumber
&=\boldsymbol{S}\partial_{\mu}\boldsymbol{L}^{-1}\boldsymbol{L}\boldsymbol{S}^{-1}
\!+\!\boldsymbol{S}\boldsymbol{\Omega}_{\mu}\boldsymbol{S}^{-1}=\\
&=\boldsymbol{S}\left(\partial_{\mu}\boldsymbol{L}^{-1}\boldsymbol{L}
\!+\!\boldsymbol{\Omega}_{\mu}\right)\boldsymbol{S}^{-1}
\end{eqnarray}
showing that the object $\partial_{\mu}\boldsymbol{L}^{-1}\boldsymbol{L}\!+\!\boldsymbol{\Omega}_{\mu}$ transforms as one spinorial matrix. Writing it as $\boldsymbol{\Omega}_{\mu}
\!-\!\boldsymbol{L}^{-1}\partial_{\mu}\boldsymbol{L}$ using (\ref{LdL}) and the known decomposition of the spinorial connection
\begin{eqnarray}
&\boldsymbol{\Omega}_{\mu}\!=\!\frac{1}{2}\Omega_{ij\mu}\boldsymbol{\sigma}^{ij}\!+\!iqA_{\mu}
\end{eqnarray}
we arrive at (\ref{P}-\ref{R}), which are therefore demonstrated to be real tensors. The $P_{\nu}$ and $R_{ij\nu}$ still contain information about gauge and frames. Hence $P_{\nu}$ such that $F_{\mu\nu}\!=\!0$ is what contains the information about gauge only whereas $R_{ij\nu}$ such that $R_{ij\mu\nu}\!=\!0$ is what contains the information about frames only. Then electrodynamics and gravity are encoded within Maxwell strength and Riemann curvature (\ref{Faraday}-\ref{Riemann}). All these objects and conditions are covariant as proven above and well known. As mentioned above, one can find non-zero $P_{\nu}$ and $R_{ij\nu}$ solutions of the conditions $F_{\mu\nu}\!=\!0$ and $R_{ij\mu\nu}\!=\!0$ respectively. An example is found in reference \cite{Fabbri:2020ypd}. What this means is that we can have a case of non-trivial background ($P_{\nu}\!\neq\!0$ and $R_{ij\nu}\!\neq\!0$) even in absence of any external force ($F_{\mu\nu}\!=\!0$ and $R_{ij\mu\nu}\!=\!0$).

This situation seems to suggest the existence of physical effects that are non-trivial albeit described by objects that cannot be determined by changes imposed through field equations with a source. To see that this is in fact the case, let us ask what could be a possible form for one candidate field equation with a source. Or within a more general approach, what is the form of one candidate field equation (for the moment regardless of the source).

In order to find in what way the Goldstone fields could be dynamically determined, we look for second-order differential field equations. Since the $R_{ij\nu}$ tensor is already first-order derivative, we only need to look for first-order derivatives of the $R_{ij\nu}$ tensor. Because we are considering field equations for $R_{ij\nu}$ with $R_{ij\mu\nu}\!=\!0$ we are considering  field equations that are not the Einstein equations. What we can do to simplify the issue is writing $R_{ij\nu}$ split as
\begin{eqnarray}
&R_{ijk}\!=\!\Pi_{ijk}\!+\!\frac{1}{3}(R_{i}\eta_{jk}\!-\!R_{j}\eta_{ik})
\!+\!\frac{1}{3}\varepsilon_{ijka}B^{a}
\end{eqnarray}
where
\begin{eqnarray}
&R_{a}\!=\!R_{ac}^{\phantom{ac}c}
\end{eqnarray}
is the trace and
\begin{eqnarray}
&B_{a}\!=\!\frac{1}{2}\varepsilon_{aijk}R^{ijk}
\end{eqnarray}
the dual of its completely antisymmetric part and with $\Pi_{ijk}$ such that $\Pi_{ia}^{\phantom{ia}a}\!=\!0$ and $\Pi_{ijk}\varepsilon^{ijka}\!=\!0$ hold. In order to find the field equations for $R_{ij\nu}$ such that $R_{ij\mu\nu}\!=\!0$ now we will have to look for field equations for each irreducible part and subject to the $R_{ij\mu\nu}\!=\!0$ constraint. As a start, we consider the last irreducible part $\Pi_{ijk}$\! for which we can immediately see that this component does not appear in the dynamics of the spinor field at all. As for the others, we must check the consistency of the field equations that have $\nabla_{a}B^{a}$ and $\nabla_{a}R^{a}$ as leading terms. Because we have to enforce the constraint given by $R_{ij\mu\nu}\!=\!0$ then
\begin{eqnarray}
&R_{\rho\alpha\mu\nu}\varepsilon^{\rho\alpha\mu\nu}\!=\!0
\end{eqnarray}
and
\begin{eqnarray}
&R_{\rho\alpha\mu\nu}g^{\rho\mu}g^{\alpha\nu}\!=\!0
\end{eqnarray}
which respectively give
\begin{eqnarray}
&\nabla_{\mu}B^{\mu}\!=\!\frac{1}{2}\varepsilon^{\alpha\sigma\mu\nu}
R_{\kappa\alpha\mu}R^{\kappa}_{\phantom{\kappa}\sigma\nu}
\end{eqnarray}
and
\begin{eqnarray}
&\nabla_{\mu}R^{\mu}\!=\!-\frac{1}{2}(\frac{1}{2}R^{\alpha\mu\nu}R_{\alpha\mu\nu}
\!+\!B^{\mu}B_{\mu}\!-\!R_{\nu}R^{\nu})
\end{eqnarray}
showing that either derivative term is reduced to an algebraic constraint, and hence the dynamical behaviour is left not determined. For $P^{a}$ a very similar argument may be used. In general, therefore, one cannot find for any of the components of the $R_{ij\nu}$ and $P_{\nu}$ tensors a differential field equation (at least, for any of the commonly-accepted definitions we have for the fundamental field equations).

This suggests that there might always be some form of dynamical under-determination for a Goldstone field like the one pertaining to the polar form of spinor fields.

This under-determination common to both Goldstone fields and hidden variables seems to point toward a link between them. Goldstone fields as hidden variables have already been used in \cite{Fabbri:2022dwj} as a way to describe correlations between a pair of opposite-spin spinorial fields.
\section{Observers}
The analysis we have done up to this point has served two purposes. One was to demonstrate that the dBB interpretation can indeed be obtained, once the polar form is used, also in the case of relativistic spinning fields, like the Dirac spinor. This led us to see that hidden variables can in fact be contextual even within a manifestly covariant environment,\! since non-local characters might appear without violating causal restrictions. The other purpose was to show that among all physical effects there may be some that are not determined by field equations. Hence, not all fields are local since not all fields are solutions of field equations, although causality has to be a property of all solutions of field equations. This leave us with some doubt about the fact that, if some fields are not solutions and some fields are solutions of field equations, then \emph{how do we fix fields that are not solutions of field equations}?

In \cite{Fabbri:2022dwj}, we presented a toy model of entangled spins, in which a pair of opposite-helicity spinors in uniform spin flip could be made to collapse simultaneously for the two fields without involving acausal processes. In short, that analysis showed that if one spinor is locked to one given helicity then the other spinor is immediately locked to the opposite helicity. However, we did not discuss in any way how the locking of the first spinor occurs. With reference to the notation of \cite{Fabbri:2022dwj}, the above statement could be re-phrased by saying that we did not discuss at all in what manner the $\omega\!\rightarrow\!0$ condition might have possibly arisen.

This problem is not new. It is in fact easy to see that it can be re-stated by asking what is the role of observers in quantum mechanics. The problem is still one of the most important and we are not going to give a solution in the following. Yet, we trust that in the toy model presented in \cite{Fabbri:2022dwj} and re-discussed above, such a problem may have a somewhat clearer mathematical formulation.
\section{Conclusion}
In this work, the Dirac spinor field theory was written, taking advantage of the polar decomposition in its manifestly covariant form, in what can then be defined as the de Broglie-Bohm formulation in relativistic version with spin. We have discussed its non-relativistic and spinless limits. And we have proven that the dBB formulation is in fact contained in it. We have then discussed the roles of contextuality and under-determination. And we have shown that the theory does provide us with objects that can be seen as hidden variables in full compatibility with causal restrictions. We have commented on observers.

Comparison between the most general relativistic version and the non-relativistic version of the dBB formulation shows that the general form is much richer, not only for the appearance of the velocity contributions. The single most important new element is the chiral angle, that is the phase difference between the two irreducible chiral projections, which encodes a form of internal dynamics that is responsible for the failure of non-relativistic limit even in the spinor rest frame. Nevertheless, for our purpose, the most illuminating feature of the general form is that the spinor field is expressed as (\ref{spinor}). Here it is clear that the relativistic form has the global $\phi e^{-iq\alpha}$ we would have in the non-relativistic form times the chiral phase $e^{-i\beta\boldsymbol{\pi}/2}$ and times a complex Lorentz transformation $\boldsymbol{\Lambda}$ which is recognized to be parametrized in terms of the well known Goldstone fields.\! We have argued that $\boldsymbol{\Lambda}$ may well be the missing element needed by Takabayasi when he attempted to find the relativistic form of the dBB formulation \cite{tr1,tr2}. In fact, after that $\boldsymbol{\Lambda}^{-1}\partial_{\mu}\boldsymbol{\Lambda}$ combines with the spin connection, the Goldstone fields become the longitudinal component of the $R_{ij\nu}$ tensor, as is clear from (\ref{R}). This expression is fundamental to ensure the full manifest covariance for the polar form of the Dirac equations (\ref{dep1}-\ref{dep2}) and these are exactly what Takabayasi would have needed to attain a relativistic version of the dBB model \cite{t}. The manifestly-covariant polar form of the Dirac equations can be written as in (\ref{ddb1}-\ref{ddb2}), which are equations giving the momentum $P_{\mu}$ by means of the quantum potentials in (\ref{Y}-\ref{Z}) and as such, they are the Hamilton-Jacobi equation for the ensemble of trajectories by construction. The expression (\ref{momentum}) is thus the guidance equation. The full mathematical setting of the dBB form of the Dirac theory is then recovered. As for the ontology, we have to assign a status to the particle. We discussed how for the Dirac theory, the spin-torsion coupling gives, if torsion is in effective approximation, a non-linear character to the field equations which then may allow solitonic solutions. These localized matter field distributions may be seen as the depiction of the particle, as de Broglie first conceived it. In this case, the hidden variables would not be the initial positions only, but also the spin orientation, and more generally all the boundary conditions entering in the structure of Goldstone fields. In \cite{Fabbri:2022dwj} we discussed how Goldstone fields may be non-local, and more in general contextual. And here we have shown that they also obey no differential field equation. Therefore, the hidden variables result to be characterized by a form of dynamical under-determination. This feature is directly linked to the presence of Goldstone fields in $\boldsymbol{\Lambda}$ and as such it is present only in the general formulation. In this sense we say that the relativistic form contains more than just the velocity when compared to the non-relativistic form.

The next natural step to follow is to enlarge this theory as to include multi-particle states. We are not even going to tackle this problem.\! Nevertheless, we are of the opinion that with the formulation presented in this paper further developments in this direction will be easier.

\

\textbf{Acknowledgments}

I wish to thank Dr. Marie-H\'{e}l\`{e}ne Genest for the useful discussions that we have had on this subject.

\end{document}